\documentclass[a4paper]{jpconf}
\pdfoutput=1
\usepackage{graphicx}
\usepackage{lineno}
\begin{document}
\title{Searches for neutrinoless double beta decay}
\linenumbers
\author{Bernhard Schwingenheuer}

\address{Max-Planck-Institut f\"ur Kernphysik, 69117 Heidelberg, Germany}

\ead{b.schwingenheuer@mpi-hd.mpg.de}

\begin{abstract}
Neutrinoless double beta decay is a lepton number violating 
process whose observation
would also establish that neutrinos are their own anti-particles. There
are many experimental efforts with a  variety of techniques.
Some (EXO, Kamland-Zen, GERDA phase I and CANDLES) started take data
in 2011 and EXO has reported the first measurement of the
half life for the double beta decay with two neutrinos of $^{136}$Xe.
The sensitivities of the different proposals are reviewed.
\end{abstract}

\section{Introduction}

For many isotopes like $^{76}$Ge $\beta$ decay is energetically
forbidden, but double beta decay ($2\nu\beta\beta$) is allowed
\begin{equation}
 (A,Z) \rightarrow (A,Z+2) + 2e^- + 2\bar{\nu_e}
\end{equation}
This was suggested
very early \cite{goeppert} and - following the idea
of Majorana that neutrinos could be their own
anti-particle \cite{majorana} -  also the possibility of 
neutrinoless
double beta decay $0\nu\beta\beta$ was anticipated 
shortly afterwards \cite{furry} (for a review see \cite{barabash,tretyak}). 
The latter case is very interesting
since lepton number is violated and it would establish
that the neutrino is its own anti-particle.
The experimental signature in this case is a line at 
the $Q_{\beta\beta}$ value of the decay if the sum
of the electron energies is plotted.

Searches for double beta decay date back to the beginning of
nuclear physics and nowadays more than a dozen large scale
experimental programs are suggested. These programs are
compared in this article and also the status of theoretical
matrix element calculations is discussed. For general
reviews the reader is refered to the literature, 
e.g.~\cite{engel,rodejohann,gomez}.

There are also other related processes like double positron decay
or double electron capture processes. While $0\nu\beta\beta$ is already a suppressed process,
the other decays are expected to
be even rarer unless there is some resonance enhancement
\cite{suhonen1,tretyak2,tgv,danevich}. 
In this article only $0\nu\beta\beta$ decay searches
are discussed.

\section{Motivation}

The observation of neutrino oscillation establishes that
these particles have mass \cite{pdg}. Since neutrinos
have no electric charge,
there is no known symmetry which forbids 
additional terms in the effective Lagrangian beside the 
standard Dirac mass term $m_D$ \cite{rodejohann}:
\begin{eqnarray}
- L_{\rm Yuk} & = & m_D \overline{\nu_L} \nu_R +  \frac{1}{2}
                m_L \overline{\nu_L}  (\nu_L)^c
              + \frac{1}{2} m_R \overline{(\nu_R)^c} \nu_R  + h.c. \\
            & = & \frac{1}{2}
  ( \overline{\nu_L},\,\, \overline{(\nu_R)^c} )
 \left( \begin{array}{cc} m_L & m_D \\ m_D & m_R \end{array} \right)
 \left( \begin{array}{c} (\nu_L)^c \\ \nu_R \end{array} \right) + h.c. 
\end{eqnarray}
The subscript $L$ stands 
for the left-handed chiral field $\nu_L = \frac{1}{2}(1-\gamma_5)\, \,\nu$ and
 $R$  for the right-handed projection 
 $\frac{1}{2}(1+\gamma_5)\,\, \nu$. The superscript $C$ 
 denotes charge conjugation.
The $m_R$ term describes therefore an  incoming neutrino and an outgoing
anti-neutrino, i.e.~lepton number is violated by 2 units.
The eigen states of the mass matrix are of the
form $(\nu+\nu^c)$. Consequently neutrinos are 
expected to be - in general - their
own anti-particles, i.e.~Majorana particles.

What is the best experimental approach to establish that 
our known neutrinos are Majorana particles? Neutrinos
(or anti-neutrinos)
are produced in charged weak current reactions and - depending on
the charge of the associated lepton -  only one chiral 
projection couples.
For example in $\beta$ decay $n\rightarrow p \, e^- \, \bar{\nu}_{e,R}$,
a right-handed anti-neutrino couples:
\begin{equation}
  \bar{\nu}_{e,R} = \bar{\nu}_e \frac{1}{2}(1+\gamma_5)
           = \sum\limits_{i=1}^{3} U_{ei} (\bar{\nu}_{i,h=+1} + 
                \frac{m_i}{E}\bar{\nu}_{i,h=-1})
\end{equation}
Here, $U$ is the PMNS mixing matrix, $\nu_i$ are the mass
eigen states with mass eigen values $m_i$, 
$E$ is the neutrino energy and $h$ stands for the
helicity of the anti-neutrino.

For a Dirac particle these anti-neutrinos can only
undergo  detection reactions like $p\, \bar{\nu}_{e,R} \rightarrow n\, e^+$.
If, on the other hand, neutrinos are Majorana particles, then
the $\nu_{i,h=-1}$ component can undergo the reaction 
 $\nu_{e,L} \, n \rightarrow p \, e^-$ with
\begin{equation}
  {\nu_{e,L}} = \frac{1}{2}(1-\gamma_5) {\nu_e} 
           = \sum\limits_{i=1}^{3} U_{ei} ({\nu_{i,h=-1}} + 
                \frac{m_i}{E} {\nu_{i,h=+1}})
\end{equation}
The rate of this  reaction\footnote{Since the charge 
of the outgoing lepton is the same as in the
production process,
$U$ and not $U^*$ enters here.}
is however suppressed by the factor $(m_i/E)^2$ which
is e.g.~$10^{-14}$ for a neutrino mass of 0.1~eV and 
a neutrino energy of 1~MeV. Thus solar neutrino
experiments for example will not be able to establish the nature
of neutrinos.


The alternative is the search for $0\nu\beta\beta$ where the
neutrino only enters as a propagator 
$\simeq m_{\beta\beta}/q^2 =\sum_i U_{ei}^2\cdot m_i/q^2 $.
The coupling strength $m_{\beta\beta}$ is called the effective Majorana
mass. Since one
mole contains a large number of nuclei, the factor
$(m_i/E)^2$ is compensated. 
For 35
isotopes double beta decay is the only possible
decay mode. The Standard Model allowed 
decay with two emitted neutrinos ($2\nu\beta\beta$)
has been observed for 11
 isotopes with half lives
between $7\cdot 10^{18}$~y  and $2\cdot 10^{21}$~y 
\cite{barabash2nu,exo}.


Part of the Heidelberg-Moscow experiment 
claims to have observed $0\nu\beta\beta$ of
$^{76}$Ge with $m_{\beta\beta} \approx 0.2-0.6$~eV \cite{klapdorplb}.
 Clearly this needs independent confirmation 
which poses another motivation for the 
experimental efforts.

\section{Experimental sensitivity}

An experiment will observe some background events $\lambda_{\rm bkg}$ which
- if this number scales by the detector mass $M$ - is given by
\begin{equation}
\label{eq:nbkg}
\lambda_{\rm bkg} = M \cdot t \cdot B \cdot \Delta E
\end{equation}
and possibly signal events 
\begin{equation}
\label{eq:nsig}
  \lambda_{\rm sig} = \ln 2 \cdot N_A \cdot \epsilon \cdot
    \eta \cdot M \cdot t /(A \cdot T_{1/2}^{0\nu}).
\end{equation}
Here  $t$ is the
measurement time, $B$ the so called background
index given typical in cnts/(keV$\cdot$kg$\cdot$y),
$\Delta E$ is the width of the search window which
depends on the experimental energy resolution,
$N_A$ is the Avogadro constant, $\epsilon$ the
signal detection efficiency, $\eta$ the mass fraction
of the $0\nu\beta\beta$ isotope, 
$A$ the molar mass of this isotope and $T_{1/2}^{0\nu}$
its half life.

\begin{table}
\caption{\label{tab:g0nu} List of most interesting $0\nu\beta\beta$
isotopes. Half lives are taken from \cite{barabash2nu,exo} while all
other numbers are from \cite{rodejohann}.}
\begin{tabular}{lccccl}
 isotope  & $G^{0\nu}$    & $Q_{\beta\beta}$ & nat.~abund. & $T_{1/2}^{2\nu}$ & experiments \\
          & $[\frac{10^{-14}}{y}$] & [keV]        &  [\%]       & [10$^{20}$ y] &  \\ \hline
 $^{48}$Ca & 6.3  & 4273.7  & 0.187  & 0.44 & CANDLES \\
 $^{76}$Ge & 0.63 & 2039.1  & 7.8   & 15 & GERDA, Majorana Demonstrator \\
 $^{82}$Se & 2.7  & 2995.5  & 9.2   & 0.92 & SuperNEMO, Lucifer \\
 $^{100}$Mo & 4.4 & 3035.0  & 9.6   & 0.07 & MOON, AMoRe \\
 $^{116}$Cd & 4.6 & 2809  & 7.6   & 0.29 & Cobra \\
 $^{130}$Te & 4.1 & 2530.3  & 34.5  & 9.1 & CUORE \\
 $^{136}$Xe & 4.3 & 2461.9  & 8.9   & 21 & EXO, Kamland-Zen, NEXT, XMASS \\
 $^{150}$Nd & 19.2 & 3367.3 & 5.6   & 0.08 &SNO+, DCBA/MTD \\
\end{tabular}
\end{table}

If $\lambda_{\rm bkg} < 1$ the experimental sensitivity
scales with $M\cdot t$ while for $\lambda_{\rm bkg} >> 1$
the e.g.~90\%\, C.L.~limit on the half life (assuming there
is no signal) is given by
\begin{equation}
\label{eq:comp}
         T_{1/2}^{0\nu}(90\% CL) > \frac{\ln 2}{1.64} \frac{N_A}{A}\,
               \epsilon\cdot \eta \cdot\sqrt{\frac{M\cdot t}{B\cdot \Delta E}}. 
\end{equation}
If systematic errors become important e.g.~if the energy resolution
is not well known or the assumption of the background shape is
not correct, then the sensitivity is reduced.

\section{Theoretical considerations}

The half life for $0\nu\beta\beta$ is given by \cite{rodejohann}
\begin{equation}
  [T_{1/2}^{0\nu}]^{-1} = 
              G_{0\nu}(Q_{\beta\beta},Z) |M_{0\nu}|^2 \frac{m_{\beta\beta}^2}{m_e^2}
\end{equation}

Here $G_{0\nu}$ is the calculable phase space factor 
(Tab.~\ref{tab:g0nu}), $m_e$
is the electron mass
and $M_{0\nu}$ is the nuclear
matrix element  whose calculation
is difficult and can only be done using approximations. 
For a review see for example \cite{engel,gomez}.

\begin{figure}
 \includegraphics[height=8cm]{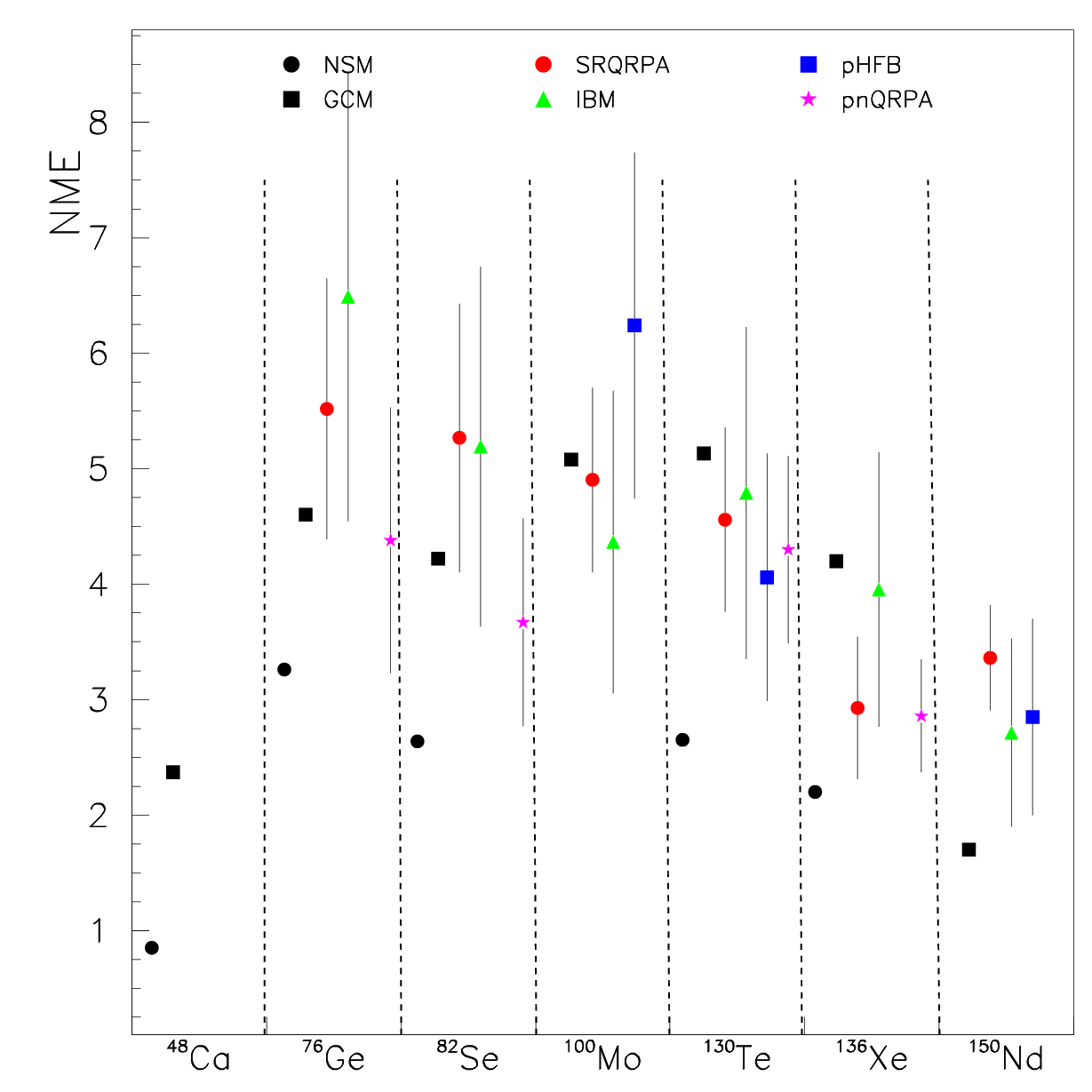}
 \includegraphics[height=8cm]{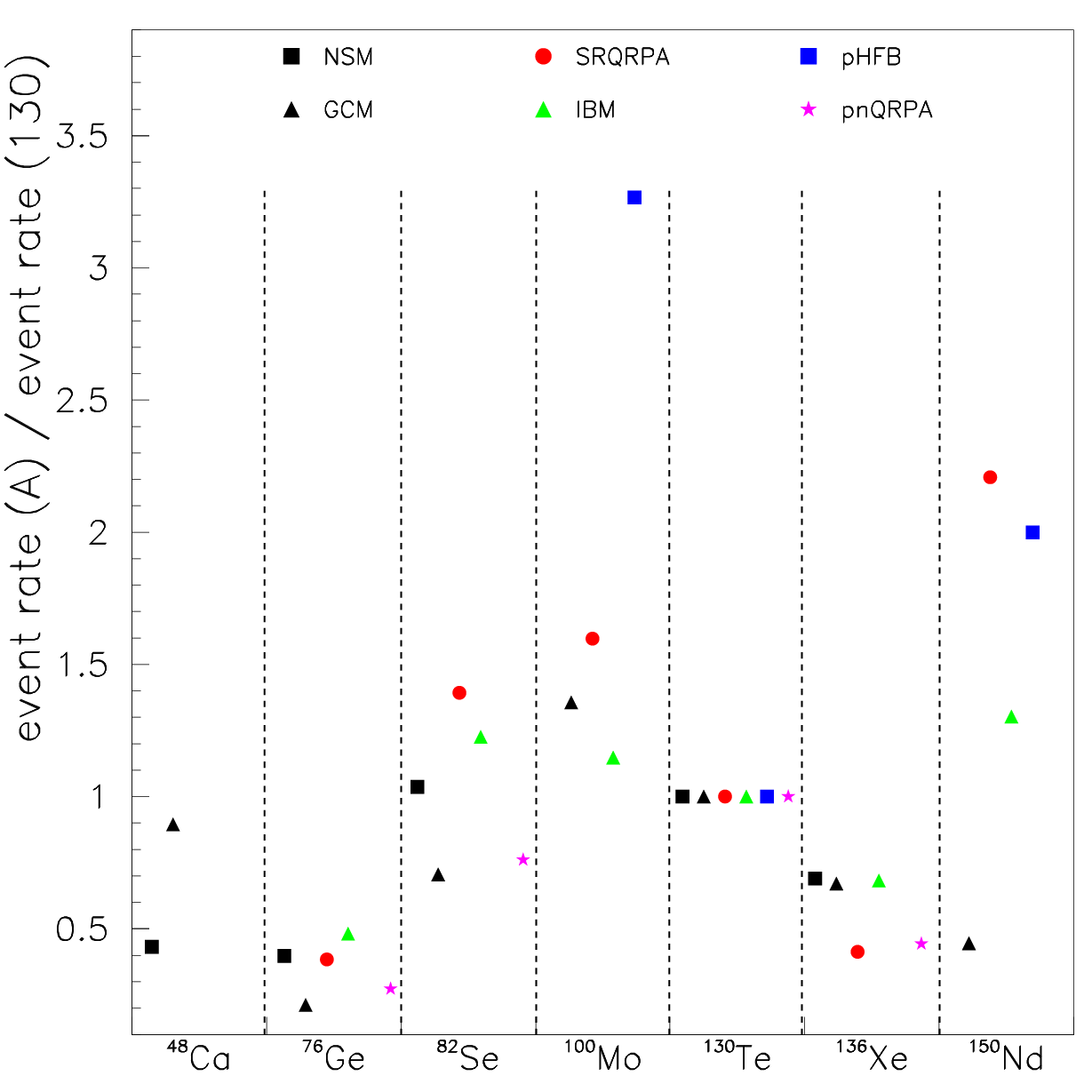}
\caption{\label{fig:nme} Left: Different calculations for nuclear matrix 
  elements for $0\nu\beta\beta$ decay for light neutrino exchange. 
  NSM = Nuclear shell model \cite{nsm,nsm2}, 
  SRQRPA = self-consistent renormalized quasi-particle random 
  phase approximation \cite{susytub,srqrpa2,srqrpa3} (matrix elements are
  scaled by 1.14 to compensate for different phase space factors), 
   pnQRPA = proton-neutron quasi particle random phase 
   approximation \cite{suhonen2},
  GCM = generating coordinate method \cite{gcm}, IBM = interacting
  boson model \cite{ibm,ibm2} (matrix elements are scaled by 1.18 to estimate the
  effect if the UCOM short range correlation instead of the Jastrow type would
  have been used \cite{gomez}), pHBF=
  projected Hartree-Fock-Bogoliubov model \cite{phfb}.
  Right: ratio of expected $0\nu\beta\beta$ events per kg target mass
  for the different models normalized to $^{130}$Te.}
\end{figure}

While the observation of $0\nu\beta\beta$  would manifest lepton number
violation and the neutrino's Majorana nature, the underlying physics 
can only be
disclosed if the observed $T_{1/2}^{0\nu}$ 
for different isotopes and possibly other
variables like the angle between the emitted electrons is compared to
theory. Consequently, there is a large interest in  nuclear matrix
element calculations and substantial progress has been made during
the last years. Traditionally, nuclear shell model (NSM) 
 and quasi particle random phase 
approximation (QRPA) calculations have been performed.
Recently new approaches like the interacting boson model (IBM),
the generating coordinate model (GCM) and the projected 
Hartree-Fock-Bogoliubov (pHFB) method have been applied. 
A discussion of these calculations is given in \cite{rodin}.

The results of these calculations are shown in Fig.~\ref{fig:nme}.
The following statements can be made concerning the status:
\begin{itemize}
\item There is no large variation for the NME 
   between the different isotopes. This might be due to the
   fact that only neighboring neutrons in a nucleus contribute
   to the decay  \cite{nsm,srqrpa2}. 
\item For the NSM, all values are systematically lower than for other methods.
      Possible reasons for this effect are discussed in the literature
      \cite{nsm,escuderos}.
\item The differences between the QRPA calculations of different
      groups are now quite small.
\item For a given isotope the calculations spread by typically a factor of 2,
      i.e.~a factor of 4 for $T_{1/2}^{0\nu}$.
\item The role of short range correlations has been studied and the UCOM
      correction has emerged as standard \cite{gomez1}. Alternatively, a self 
      consistent implementation
      was first applied to SRQRPA \cite{srqrpa2}
      and later to other methods \cite{nsm2,phfb} and resulted in
      small changes. 
\item Experimental input can have a large shift of the result. For example
      charge exchange reaction measurements of $^{150}$Nd($^3$He,t)
      and  $^{150}$Sm(t,$^3$He) \cite{zegers}
      result in a quenching factor of 0.75 for the
      $g_A$ coupling and hence a reduction of the matrix element by 25\%
      for $^{150}$Nd \cite{srqrpa3}. In this calculation, deformation
      was treated for the first time in a QRPA calculation.

      For $^{76}$Ge and $^{76}$Se,  the proton and neutron valence orbital 
      occupancies have been measured \cite{schiffer,schiffer2}. 
      If the models are adjusted to reproduce
      these values, the NSM result increases by 15\% \cite{nsm2} 
      while the QRPA
      results are reduced by about 20\% \cite{simkovic,suhonen3}. 
      Hence the difference between NSM and QRPA  becomes  half as large.
\end{itemize}

The calculations are performed for the standard 
light neutrino exchange but results
for other mechanisms like SUSY particle exchange 
are also available \cite{hirsch,susytub}.

In order to see whether some isotopes are better suited
for $0\nu\beta\beta$ decay searches from a theoretical 
point of view,
the number of expected decays per isotope mass can be compared.
This value includes the phase space factor, the matrix element and
the mass number $A$. For the comparison it is sufficient to look at
the ratio of decay rates and in this case, some of the systematic effects
of the matrix element calculations cancel since there are
typically correlations among the isotopes for a given method.
 The right hand plot of Fig~\ref{fig:nme}
shows these ratios for the different models normalized to the
decay rate of $^{130}$Te. One sees that $^{76}$Ge is less favorable.
The expected decays per kg  vary between 20\% and 50\% of the
rate of $^{130}$Te. In other words: if all experimental parameters
were the same then one would need a factor of $\approx$3 more target mass
in a $^{76}$Ge experiment to have the same sensitivity.
In reality this is not the case, i.e.~the superior energy
resolution of Ge detectors compensates this effect.  

\section{Comparison of experiments}

The experiments searching for $0\nu\beta\beta$ decay
use a large variety of  detection
mechanisms and background reduction methods, see Tab.~\ref{tab:list}.
The current status of almost all of them is described in
these proceedings. Therefore a more detailed discussion 
is omitted here. Instead the key performance numbers are
taken for a comparison of the sensitivities of some experiments.

\begin{table}
\caption{\label{tab:list} Selection of $0\nu\beta\beta$ experiments.}
\begin{tabular}{lcccccc}
experiment & isotope & mass [kg] & method & location & time & ref.  \\ \hline
\multicolumn{6}{c}{past experiments } \\ 
Heidelberg-Ms. & $^{76}$Ge & 11 & ionization & LNGS & -2003 &\cite{klapdorplb}  \\
Cuoricino & $^{130}$Te & 11 & bolometer & LNGS  & -2008 & \cite{cuoricino} \\
NEMO-3 & $^{100}$Mo,$^{82}$Se & 7,1 & track.+calorim. & Modane & -2011 & \cite{nemo3}\\ \hline
\multicolumn{6}{c}{ current experiments } \\
EXO    & $^{136}$Xe & 175 & liquid TPC & WIPP & 2011- & \cite{exo} \\
Kamland-Zen & $^{136}$Xe & 330 & liquid scintil. & Kamioka & 2011- & \cite{kamland} \\
GERDA-I/II   & $^{76}$Ge  & 17/35  & ionization      & LNGS    & 2011-/13 & \cite{gerda} \\
CANDLES     & $^{48}$Ca  & 0.35 & scint.~crystal & Oto Cosmo & 2011- & \cite{candles} \\ \hline
\multicolumn{6}{c}{funded experiments} \\
NEXT        & $^{136}$Xe & 100 & gas TPC & Canfrac & 2014 & \cite{next} \\
Cuore0/Cuore & $^{130}$Te & 10/200 & bolometer & LNGS & 2012/14 & \cite{cuore} \\
Majorana Demo. & $^{76}$Ge & 30 & ionization & SUSEL & 2014 & \cite{john} \\
SNO+         & $^{150}$Nd & 44 & liquid scint. & Sudbury & 2014 & \cite{sno} \\ \hline
\multicolumn{6}{c}{proposal, proto-typing} \\
SuperNEMO   & $^{82}$Se  & 7/100-200 & track.+calorim. & Modane & 2014/- & \cite{supern}\\
Cobra       & $^{116}$Cd &       & solid TPC & LNGS & &\cite{cobra}  \\
Lucifer     & $^{82}$Se &      & bolom.+scint. & LNGS & & \cite{lucifer} \\
DCBA/MTD    & $^{150}$Nd & 32 & tracking       &      & & \cite{dcba} \\
MOON        & $^{82}$Se,$^{100}$Mo & 30-480  & track.+scint. & & & \cite{moon}  \\
XMASS       & $^{136}$Xe &  & liquid scint. & Kamioka &  & \cite{xmass} \\
AMoRE      & $^{100}$Mo & 100  & bolom.+scint. & YangYang & & \cite{amore} \\ 
Cd exp.    & $^{116}$Cd &      & scint. & & & \cite{cdexp} \\ \hline
\end{tabular}
\end{table}

Since experiments use different isotopes a relative scaling factor for the
different matrix elements and phase spaces has to be applied. This
factor can be estimated using Fig.~\ref{fig:nme}. The values used 
here are $f_A({\rm Ge})=0.35$, $f_A({\rm Se})=1.1$,
$f_A({\rm Mo})=1.6$, $f_A({\rm Te})=1$,
$f_A({\rm Xe})=0.55$ and $f_A({\rm Nd})=2.2$.

If  the number of background events is
large, equation~\ref{eq:comp} can be used to estimate the experimental
sensitivity. 
A relative figure-of-merit can  then be defined as
\begin{equation}
 {\rm FOM} = f_A\cdot\epsilon\cdot\eta\cdot\sqrt{\frac{M}{B\cdot\Delta E}}
\end{equation}
One can call this the ``ultimate'' relative sensitivity of an experiment.
Tab.~\ref{tab:comp} lists the performance numbers and the figure-of-merit.
For running (and past) experiments  like EXO and GERDA-I the current achieved values
are used which might improve with time while for the others the anticipated
performance numbers are taken.\footnote{A fiducial volume cut will reduce the
active mass. Depending on whether  the background index is 
normalized to
the total mass or to the fiducial mass, the efficiency $\epsilon$ has
to go under the square root or not. The meaning of $B$
 is not always clearly defined in the literature. Here the normalization to the
total mass is assumed.}

\begin{figure}
\begin{center}
 \includegraphics[height=8cm]{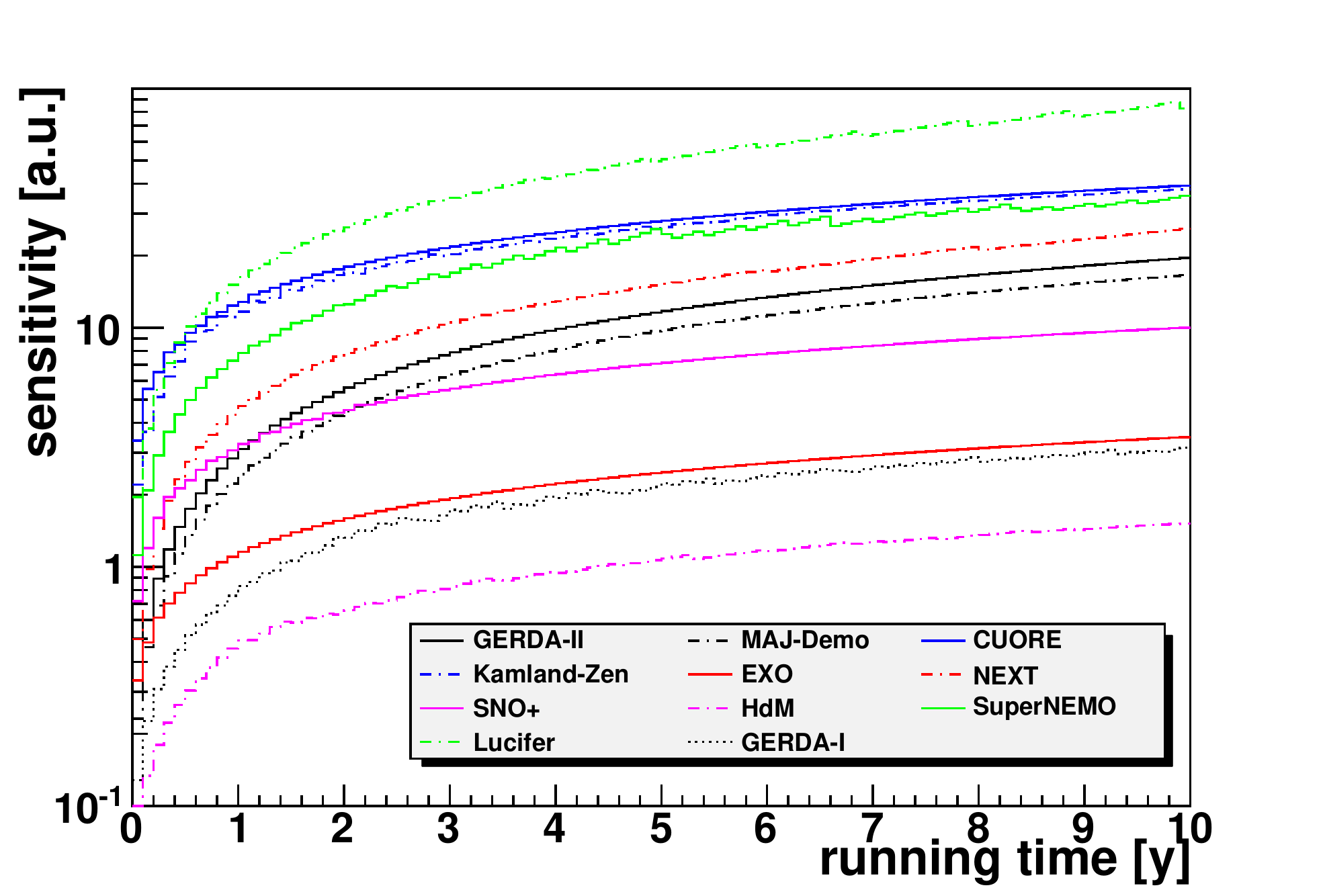}
\end{center}
\caption{\label{fig:t12} Relative experimental sensitivity for
$0\nu\beta\beta$ life time limit versus running time.}
\end{figure}

Alternatively, the (relative) sensitivity vs.~time can be estimated
from equation \ref{eq:nsig} by
\begin{equation}
    \hat{T}_{1/2}^{0\nu} > \frac{f_A \cdot \epsilon \cdot \eta \cdot M \cdot t}
              {\Psi (B\cdot \Delta E \cdot M \cdot t)} 
\end{equation}
Here $\Psi (\lambda_{\rm bkg})$ is the ``average'' 90\%C.L.~upper 
limit of the number of signal
events for $\lambda_{\rm bkg}$ background events calculated according to the
method discussed in \cite{gomez1}. The result is shown in Fig~\ref{fig:t12}.
Here all experiments are assumed to start at time 0.

\begin{table}
\caption{\label{tab:comp} Comparison of figure-of-merits (FOM)
for the  case  of large number of  background events (``ultimate
sensitivity'').
$f_A$ is the scale factor for
a given isotope taken from Fig.~\ref{fig:nme}(right), and
$\Delta E$ is the energy window which is taken here to be 1(2) full width half maximum
for experiments with $>0.5\%$ ($<0.5\%$) resolution. Note that the efficiency is
reduced by 0.7 if $\Delta E = $~1$\cdot$FHWM.
FOM is defined in the text.}
\begin{centering}
\begin{tabular}{lccccccc} \hline
experiment & mass & $f_A$ & background & $\Delta E$ & efficiency & enrichment &  FOM \\
           &[kg]& & [$\frac{{\rm cnt}}{{\rm keV}\cdot{\rm kg}\cdot{\rm y}}$] & [keV] & &  & \\ \hline
Hd-Moscow  & 11   & 0.35   & 0.12      & 8    &  0.8      & 0.86    & 0.8  \\
Cuoricino  & 41   &  1     & 0.16      & 12   &  0.9       & 0.27  &  1.1  \\
NEMO-3     & 6.9  & 1.6    & 0.002     & 240  &  0.18      & 0.9    & 1.0  \\ \hline
EXO        & 175  & 0.55   & 0.004     & 260  &  0.33      & 0.81   & 1.9 \\
Kamland-Zen& 330  & 0.55   & 0.0002    & 250  &  0.5       & 0.9    & 20  \\
GERDA-I    &  15  & 0.35   & 0.03      & 10   &  0.8       & 0.86   & 1.7  \\ \hline
GERDA-II   &  30  & 0.35   & 0.001     &  6   &  0.8       & 0.88   & 17 \\
Major.-Dem.  & 20   & 0.35   & 0.001   &  6   &  0.9       & 0.9    & 16 \\
CUORE      & 750  & 1      & 0.01      & 10   &  0.9       & 0.27   & 21  \\
SNO+       & 800  & 2.2    & 0.0002    & 230  &  0.33      & 0.056  &  5.4 \\
NEXT       & 100  & 0.55   & 0.0002    & 25   &  0.25      & 0.9    &  18  \\
SuperNEMO  & 100  & 1.1    & 0.0002    & 120  & 0.3        & 0.9    &  19  \\
Lucifer    & 100  & 1.1    & 0.001     & 10   &  0.9       & 0.5    &  50 \\ \hline
\end{tabular}
\end{centering}
\end{table}

A few comments should be made concerning the 
interpretation of Tab.~\ref{tab:comp} and Fig.~\ref{fig:t12}.
\begin{itemize}
\item If one takes the spread of the data points in Fig.~\ref{fig:nme}
      the factor $f_A$ has a $\approx 20$\% uncertainty.
\item The $2\nu\beta\beta$ background is irreducible and can only
      be avoided with an energy resolution $\sigma < 1-2$\% at
      $Q_{\beta\beta}$. This requirement depends of course strongly
      on $T_{1/2}^{2\nu}$ which varies by a factor of 300 for
      the isotopes considered. For some experiments this background is not
      fully taken into account for the background index.
\item All sensitivities given are the scales for $0\nu\beta\beta$ 
      discovery. To get relative sensitivities for
      $m_{\beta\beta}$ the square root has to be taken.
\item Of the running experiments, Kamland-Zen should have the
      largest potential. This is impressive if one takes into
      account that it was not specially built
      for this physics.
\item Germanium experiments can be very competitive despite
      the fact that the phase space factor is so small. 
      Especially if a positive signal will be claimed,
      a narrow peak at $Q_{\beta\beta}$ will be more convincing than a broad
      shoulder.
\item The Lucifer approach with 100~kg is very competitive
      even in comparison to a ton scale Xe experiment like Kamland-Zen
      or NEXT.
\item Systematic effects like the precision of the energy resolution
      or the background shape are not taken into account.
\end{itemize}

In case the neutrino masses are ordered in the inverted 
 mass hierarchy, a lower bound of
about 15~meV for $m_{\beta\beta}$ can be calculated using
the current parameters from neutrino oscillation
experiments. For $^{76}$Ge this corresponds to half lives
of $5-20\cdot 10^{27}$~years. These values should be compared
to the expected sensitivity of GERDA-II or the Majorana Demonstrator
of about 1.5$\cdot 10^{26}$~y.  This demonstrates that
exploring the entire mass band of the inverted hierarchy is a
long term enterprise. With the numbers in Tab.~\ref{tab:comp}
and a mass of 1000 kg, the required time for $5\cdot 10^{27}$~y
is 13 years while a Lucifer like experiment would need to
run for half the time.

\section{Summary}

Neutrinoless double beta decay is the best experimentally accessible method
to test whether neutrinos are Majorana particles. This decay violates
lepton number and is therefore on equal footing to proton decay searches.
The motivation for several large efforts in this field is therefore obvious.

For a long time, the Heidelberg-Moscow experiment has dominated the field and
its claim of a $0\nu\beta\beta$ signal has not been scrutinized since 2001.
In 2011,  EXO, Kamland-Zen, CANDLES and GERDA-I started to take data. All but
CANDLES 
 are more sensitive than Heidelberg-Moscow and especially
Kamland-Zen is expected to answer this question in the next 12 months.
EXO has already reported a first time measurement of 
$T_{1/2}^{2\nu} (^{136}{\rm Xe}) = 2.11\pm0.04({\rm stat})\pm 0.21({\rm syst})
\cdot 10^{21}$y which is considerably lower than previous limits \cite{exo}.

Beyond this next step, experiments want to explore the $m_{\beta\beta}$
region for the inverted neutrino mass hierarchy. This will eventually require ton scale
experiments. Which of the proposed solutions will be built is open
at the moment.

\section*{References}
\bibliography{schwingenheuer_dbd}

\end{document}